\documentclass[preprint,showpacs,preprintnumbers]{revtex4}
\usepackage{graphicx}
\usepackage{dcolumn}
\usepackage{bm}

\def\lesssim{\ \raise.3ex\hbox{$<$}\kern-0.8em\lower.7ex\hbox{$\sim$}\ }
\def\gesim{\ \raise.3ex\hbox{$>$}\kern-0.8em\lower.7ex\hbox{$\sim$}\ }
\font\scripti=cmmi7
\font\scriptscripti=cmmi5
\def\sib#1{\setbox0 = \hbox{\scripti #1}
  \kern-.02em\copy0\kern-\wd0
  \kern.04em\box0} 
\def\ssib#1{\setbox0 = \hbox{\scriptscripti #1}
  \kern-.02em\copy0\kern-\wd0
  \kern.04em\box0} 
\font\tenib=cmmib10 
\skewchar\tenib='177 \skewchar\tenib='177 \skewchar\tenib='177
\def\pbold#1{\setbox0 = \hbox{$ #1 $}
  \kern-.022em\copy0\kern-\wd0
  \kern.011em\copy0\kern-\wd0
  \kern.011em\copy0\kern-\wd0
  \kern.011em\copy0\kern-\wd0
  \kern.011em\box0} 
\begin{document}
\title{BCS-BEC crossover in a gas of Fermi atoms with a ${\bf p}$-wave Feshbach resonance}
\author{Y. Ohashi$^{1,2}$}
\affiliation{$^1$Institute of Physics, University of Tsukuba, Tsukuba,
  Ibaraki 305, Japan,\\  $^2$Department of Physics, University of Toronto, Toronto,
  Ontario, Canada M5S 1A7}
\date{\today}
\begin{abstract}
We investigate unconventional superfluidity in a gas of Fermi atoms with an anisotropic
$p$-wave Feshbach resonance. Including the $p$-wave Feshbach resonance as well 
as the associated three kinds of 
quasi-molecules with finite orbital angular momenta $L_z=\pm1,0$, 
we calculate the transition temperature of the superfluid phase. As one passes through
the $p$-wave Feshbach resonance, we find the usual BCS-BEC crossover phenomenon. The $p$-wave BCS state continuously changes into the BEC of bound molecules with $L=1$.
Our calculation includes the effect of fluctuations associated with Cooper-pairs and
molecules which are not Bose-condensed.
\end{abstract}
\maketitle
%
\par
The search for $p$-wave superfluidity is the next big challenge in a trapped Fermi gas, after the discovery of $s$-wave superfluidity in $^{40}$K and $^6$Li\cite{Jin,Bartenstein,Zwierlein,Kinast,Bourdel}. Recently, $p$-wave Feshbach resonances have been observed\cite{Regal,Salomon,Schunk}. The discovery of $p$-wave superfluidity will be the first realization of pseudo-spin {\it triplet} superfluidity in a Fermi atomic gas, quite different from the recently discovered {\it singlet} $s$-wave superfluidity\cite{Jin,Bartenstein,Zwierlein,Kinast,Bourdel}. 
Since the pairing interaction associated with a Feshbach resonance can be tuned by varying the threshold energy $2\nu$ of the resonance (see below), one can probe the $p$-wave BCS-BEC crossover. The superfluidity will continuously change from $p$-wave BCS-type to a BEC-type of bound molecules with a finite angular momentum $L=1$, as one passes through the Feshbach resonance. 
\par
As a useful first step, we calculate the superfluid phase transition temperature $T_{\rm c}$ over the entire $p$-wave BCS-BEC crossover regime.
We explicitly include the $p$-wave Feshbach resonance and associated molecules with three values of $L_z=\pm 1,0$. To describe the BCS-BEC crossover\cite{Nozieres,Ohashi}, it is necessary to include the fluctuations in the three $p$-wave Cooper-channels and their coupling due to the strong pairing interaction associated with the Feshbach resonance. Our work is a generalization of Ref. \cite{Ohashi}.
We consider both a single-component Fermi gas (where a Feshbach resonance occurs in the same hyperfine state) as well as a two-component Fermi gas (where a Feshbach resonance occurs between different hyperfine states). The Feshbach resonances in both cases have been recently observed\cite{Regal,Salomon,Schunk}. We deal with both a narrow and a broad Feshbach resonance. Although we mainly consider a uniform gas in this letter, 
we discuss $T_{\rm c}$ in a trapped gas in the BEC limit. 
\par
$p$-wave superfluidity in trapped Fermi gases was discussed in the BCS regime at $T_{\rm c}$\cite{Stoof,Bohn}. Very recently, the $p$-wave gap equation for the order parameter of the superfluid phase in the crossover region was solved in Ref. \cite{Ho} at $T=0$. An attractive interaction in the $L\ne 0$ partial wave channel was considered. An interesting phase transition in a two-dimensional Fermi gas has also been predicted\cite{Melo}. 
In contrast to these recent papers, our starting point explicitly introduces the molecules which form as a result of the Feshbach resonance. We thus emphasize the physical nature of the $p$-wave bound states which form the Bose condensate in the crossover region. We also remark that a $p$-wave pairing mechanism has been proposed using the dipole interaction\cite{Baranov}.
\par
We extend the coupled fermion-boson (CFB) model for a $s$-wave Feshbach resonance\cite{Timmermans,Holland,Ohashi} to a $p$-wave one, 
\begin{eqnarray}
H
&=&
\sum_{\bf p}\varepsilon_{\bf p}c^\dagger_{\bf p}c_{\bf p}
+\sum_{{\bf q},j}[\varepsilon^B_{{\bf q}}+2\nu] 
b_{{\bf q},j}^\dagger b_{{\bf q},j}
\nonumber
\\
&-&
{U \over 2}\sum_{{\bf p},{\bf p}',{\bf q},j}
{\bf p}\cdot{\bf p}'
c_{{\bf p}+{\bf q}/2}^\dagger
c_{-{\bf p}+{\bf q}/2}^\dagger
c_{-{\bf p}'+{\bf q}/2}
c_{{\bf p}'+{\bf q}/2}
\nonumber
\\
&+&
{g_{\rm r} \over \sqrt{2}}
\sum_{{\bf p},{\bf q},j}p_j
\Bigl[
b_{{\bf q},j} 
c_{{\bf p}+{\bf q}/2}^\dagger
c_{-{\bf p}+{\bf q}/2}^\dagger
+h.c.
\Bigr].
\label{eq.1}
\end{eqnarray}
Here $c_{\bf p}^\dagger$ is a creation operator of a Fermi atom with the kinetic energy $\varepsilon_{\bf p}\equiv p^2/2m$. $b_{{\bf q},j}$ describe three kinds of molecular bosons (labelled by $j=x,y,z$), all with the center of mass momentum ${\bf q}$, associated with the $p$-wave Feshbach resonance. 
The threshold energy $2\nu$ in the molecular kinetic energy $\varepsilon^B_{\bf q}+2\nu\equiv q/2M+2\nu$ is independent of $j$, due to the spherical symmetry of the system we are considering. In the last term, $g_{\rm r}$ is the coupling constant of a $p$-wave Feshbach resonance, with $p_j$ characterizing the $p$-wave symmetry\cite{Voll}. 
\par
The Feshbach resonance term in (\ref{eq.1}) is obtained from a more general Hamiltonian
$H_{\rm F.R.}\equiv \int d{\bf r}d{\bf r}' [g_{\rm r}({\bf r}-{\bf r}')\Phi({\bf r},{\bf r}')\Psi^\dagger({\bf r})\Psi^\dagger({\bf r}')+h.c.]$. Here $\Psi({\bf r})\equiv\sum_{\bf p}e^{i{\bf p}\cdot{\bf r}}c_{\bf p}$ is a fermion field 
operator, and 
\begin{eqnarray}
\Phi({\bf r},{\bf r}')\equiv \sum_{\bf q} e^{i{\bf q}\cdot{\bf R}}
\sum_{n,L,L_z}u_{nL}({\tilde r})Y_{L,L_z}(\theta,\phi) b_{{\bf q},n,L,L_z}
\label{eq.1b}
\end{eqnarray}
describes molecules with center of mass ${\bf R}\equiv({\bf r}+{\bf r}')/2$, and relative coordinate ${\tilde {\bf r}}\equiv{\bf r}-{\bf r}'$. $b_{{\bf q},n,L,L_z}$ is an annihilation operator of a bound molecular state, described by the eigenfunction $u_{nL}({\tilde r})Y_{L,L_z}(\theta,\phi)$. The last term in (\ref{eq.1}) is obtained when we retain the terms in $\Psi({\bf r})$ to leading order in ${\bf p}$, in the $L=1$ channel 
($b_{{\bf q}L_z}\equiv b_{{\bf q},n,L=1,L_z}$) for a Feshbach resonance state specified by a radial quantum number $n$. 
We note that $p_x\propto Y_{11}+Y_{1,-1}$, $p_y\propto Y_{11}-Y_{1,-1}$ and $p_z\propto Y_{1,0}$. Thus the molecular operators $b_{{\bf q},j}$ in (\ref{eq.1}) are related to $b_{{\bf q},L_z}$, with azimuthal angular momentum components $L_z=\pm1, 0$, as follows $(b_{{\bf q},x},b_{{\bf q},y},b_{{\bf q},z})=
({1 \over \sqrt{2}}[b_{{\bf q},1}+b_{{\bf q},-1}], {i \over \sqrt{2}}[b_{{\bf q},1}-b_{{\bf q},-1}], b_{{\bf q},0})$. 
Equation (\ref{eq.1}) also includes a non-resonant $p$-wave interaction $U$\cite{Voll}, which we take to be attractive ($-U<0$).
\par
In the $p$-wave Feshbach resonance, since two Fermi atoms form one of the three kinds of quasi-molecular bosons described by $b_{{\bf q},j}^\dagger$ ($j=x,y,z$) and this bound state can dissociate into two Fermi atoms, we take $M=2m$ and impose the conservation of the total number of Fermi atoms as $N=N_{\rm F}+2\sum_{j=x,y,z}N_{\rm B}^j$. Here $N_{\rm F}$ is the number of Fermi atoms and $N_{\rm B}^j$ is the number of Bose molecules in the $j$-th component. This constraint can be actually absorbed into (\ref{eq.1}) by considering the grand-canonical Hamiltonian $H\equiv H-\mu N$. The resulting Hamiltonian has the same form as (\ref{eq.1}), where $\varepsilon_{\bf p}$ and $\varepsilon^B_{\bf q}$ are replaced by $\xi_{\bf p}\equiv\varepsilon_{\rm p}-\mu$ and $\xi^B_{\bf q}\equiv\varepsilon^B_{\bf q}-2\mu$, respectively.
\par
The superfluid phase is characterized by three anisotropic $p$-wave Cooper-pairs $\Delta_j({\bf p})\equiv U\sum_{{\bf p}'}p_jp_j'\langle c_{-{\bf p}'}c_{{\bf p}'}\rangle$ and three molecular BEC order parameters $\phi_j\equiv\langle b_{{\bf q}=0,j}\rangle$ ($j=x,y,z$). In the equilibrium state, these are related to each other through the identity\cite{Ohashi}
$p_j\phi_j=-{g_{\rm r} \over \sqrt{2}U}{1 \over 2\nu-2\mu}\Delta_j({\bf p})$. 
The single-particle excitations have the BCS spectrum $E_{\bf p}=\sqrt{\xi_{\bf p}^2+|\sum_j{\tilde \Delta}_j({\bf p})|^2}$ with the composite order parameter, given by
${\tilde \Delta}_j({\bf p})\equiv\Delta_j({\bf p})-\sqrt{2}g_{\rm r}p_j\phi_j$. 
The angular dependence of ${\tilde \Delta}_j({\bf p})$ is proportional to $p_j$. This composite order parameter is self-consistently determined 
by the BCS gap equation, 
$1={1 \over 3}U_{\rm eff}\sum_{\bf p}{p^2 \over 2E_{\rm p}}\tanh{E_{\bf p} \over 2T}$,
where the factor $p^2/3$ comes from the angular integration of $p_j^2$. The effective pairing interaction $U_{\rm eff}\equiv U+g_{\rm r}^2/(2\nu-2\mu)$ includes the effect of Feshbach resonance\cite{Ohashi}. In the weak-coupling or BCS regime, $\mu\simeq\varepsilon_{\rm F}$ (where $\varepsilon_{\rm F}$ is the Fermi energy).
\par
The analogous $p$-wave CFB model for a two-component Fermi gas ($\equiv\uparrow,\downarrow$) is described by
\begin{eqnarray}
H
&=&
\sum_{{\bf p},\sigma}\varepsilon_{\bf p}c^\dagger_{{\bf p}\sigma}c_{{\bf p}\sigma}
+\sum_{{\bf q},j}[\varepsilon^B_{{\bf q}}+2\nu] b_{{\bf q},j}^\dagger b_{{\bf q},j}
\nonumber
\\
&-&
U\sum_{{\bf p},{\bf p}',{\bf q},j}
{\bf p}\cdot{\bf p}'
c_{{{\bf p}+{\bf q}/2}\uparrow}^\dagger
c_{{-{\bf p}+{\bf q}/2}\downarrow}^\dagger
c_{{-{\bf p}'+{\bf q}/2}\downarrow}
c_{{{\bf p}'+{\bf q}/2}\uparrow}
\nonumber
\\
&+&
g_{\rm r}
\sum_{{\bf p},{\bf q},j}
p_j
\Bigl[
b_{{\bf q},j} 
c_{{{\bf p}+{\bf q}/2}\uparrow}^\dagger
c_{{-{\bf p}+{\bf q}/2}\downarrow}^\dagger
+h.c.
\Bigr].
\label{eq.5}
\end{eqnarray}
In a mean field pairing approximation, we again obtain the same single-particle excitations $E_{\bf p}$ and the gap equation as those in the single-component case. 
The Cooper-pair order parameter is $\Delta_j({\bf p})\equiv U\sum_{{\bf p}'}p_jp_j'\langle c_{-{\bf p}'\downarrow}c_{{\bf p}'\uparrow}\rangle$. 
\par
We now present the $p$-wave strong-coupling theory at $T_{\rm c}$ for the single-component (spin polarized) model defined in (\ref{eq.1}). 
The discussion is easily extended to the two-component case. The equation for $T_{\rm c}$ is obtained by employing the Thouless criterion\cite{Nozieres,Ohashi}, the temperature when the particle-particle scattering matrix first develops a pole at $\omega={\bf q}=0$.
In the $t$-matrix approximation, the $p$-wave scattering matrix has the form ${\tilde \Gamma}_{ij}({\bf p},{\bf p}',{\bf q},\omega)=p_i\Gamma_{ij}({\bf q},\omega)p_j$, which is shown diagrammatically in Fig. 1(a). In this figure, the first and the second lines, respectively, describe the effects of non-resonant interaction $U$ and the $p$-wave Feshbach resonance, that give ${\hat \Gamma}({\bf q},\omega)\equiv\{\Gamma_{ij}\}=-[1-U_{\rm eff}({\bf q},\omega){\hat \Pi}({\bf q},\omega)]^{-1}U_{\rm eff}({\bf q},\omega)$ ($i,j=x,y,z$). Here, $U_{\rm eff}({\bf q},\omega)\equiv U-g_{\rm r}^2D_0({\bf q},\omega)$ is an atom-atom interaction including dynamical effects described by the bare molecular Bose propagator $D_0^{-1}({\bf q},\omega)\equiv \omega+i\delta-[\xi^B_{\bf q}+2\nu]$. The correlation functions ${\hat \Pi}\equiv\{\Pi_{ij}\}$ are obtained from the analytic continuation $i\nu_n\to\omega+i\delta$ of the two-particle thermal Green's function,
\begin{eqnarray}
\Pi_{ij}({\bf q},i\nu_n)
\equiv
{1 \over \beta}
\sum_{\bf p}
p_ip_j
{
1-f(\xi_{{\bf p}+{\bf q}/2})-f(\xi_{{\bf p}-{\bf q}/2})
\over
\xi_{{\bf p}+{\bf q}/2}+\xi_{{\bf p}-{\bf q}/2}-i\nu_n
},
\label{eq.6}
\end{eqnarray}
where $f(\varepsilon)$ is the Fermi distribution function.
The diagonal components $\Pi_{ii}$ ($i=x,y,z$) describe superfluid fluctuations in the $i$-th Cooper-channel, while the off-diagonal components give the coupling of fluctuations in {\it different} channels. Noting that $\Pi_{i\ne j}(0,0)=0$ in our approximation, the Thouless criterion gives the equation for $T_{\rm c}$ as 
\begin{equation}
1=U_{\rm eff}\Pi_{ii}(0,0)={1 \over 3}U_{\rm eff}
\sum_{\bf p}{p^2 \over 2(\varepsilon_{\bf p}-\mu)}
\tanh{\varepsilon_{\bf p}-\mu \over 2T}.
\label{eq.7}
\end{equation}
This has the same form as the mean-field gap equation at $T=T_{\rm c}$, with ${\tilde \Delta}_j({\bf p})\to 0$. However, the chemical potential $\mu$ in (\ref{eq.7}) can be quite different from the Fermi energy $\varepsilon_{\rm F}$ in the crossover regime\cite{Ohashi,Nozieres,Eagles,Leggett}, and one needs an additional equation to determine $\mu$. 
\par
The chemical potential $\mu$ is determined from the equation for the number of atoms $N$, which is calculated from the thermodynamic potential $\Omega$ using the formula 
$N=-{\partial \Omega \over \partial \mu}$\cite{Nozieres,Ohashi}. Figure 1(b) shows the fluctuation correction to $\Omega$, 
where the diagrams on the left and right describe fluctuations in the $p$-wave Cooper-channels and the Feshbach resonance, respectively. Summing up these diagrams, we obtain
\begin{eqnarray}
N
&=&N_{\rm F}
-{1 \over \beta}
\sum_{{\bf q},i\nu_n}
{\partial \over \partial \mu}
tr
\Bigl[
\log{\hat D}({\bf q},i\nu_n)
\Bigr]
\nonumber
\\
&-&
{1 \over \beta}
\sum_{{\bf q},i\nu_n}
{\partial \over \partial \mu}
tr
\Bigl[
\log
[1-U{\hat \Pi}({\bf q},i\nu_n)]
\Bigr]
\nonumber
\\
&\equiv& N_{\rm F}+2N_{\rm B}+2N_{\rm C},
\label{eq.8}
\end{eqnarray}
where the trace is taken over the $L=1$ space ($j=x,y,z$). $N_{\rm F}\equiv\sum_{\bf p}f(\xi_{\bf p})$ is the number of free Fermi atoms. $N_{\rm B}$ is the number of Feshbach molecules, given as the poles of the renormalized (matrix) molecular Bose Green's function ${\hat D}^{-1}({\bf q},i\nu_n)\equiv i\nu_n-(\xi^B_{\bf q}+2\nu)-{\hat \Sigma}({\bf q})$. The molecular self-energy ${\hat \Sigma}({\bf q},i\nu_n)\equiv-g_{\rm r}^2{\hat \Pi}/(1-U{\hat \Pi})$ describes the fluctuation effects in the $p$-wave Cooper-channels. As in Refs. \cite{Nozieres,Ohashi}, $N_{\rm C}$ can be interpreted as the contribution of preformed $p$-wave Cooper-pairs as well as particle-particle scattering states. We note that superfluid fluctuations in the three $p$-wave Cooper-channels are strongly coupled to one another in (\ref{eq.8}) through $\Pi_{ij}$ ($i\ne j$).
Equations (\ref{eq.7}) and (\ref{eq.8}) are the basic coupled equations 
describing $T_{\rm c}$ of a uniform $p$-wave superfluid over the
entire BCS-BEC crossover. The same equations are obtained in the two-component case in (\ref{eq.5}), with $N_{\rm F}$ replaced by $N_{\rm F}=2\sum_{\bf p}f(\xi_{\bf p})$, reflecting the two Fermi hyperfine states.
\par
Figure 2 shows the self-consistent numerical 
solutions of (\ref{eq.7}) and (\ref{eq.8}). 
In this figure, we have introduced a $p$-wave scattering length $a_p$ for the renormalized interaction $U^R_{\rm eff}$ which occurs in the gap equation when written in a cutoff-independent way\cite{Ohashi2}.
This is defined\cite{Stoof,Bohn} by $-4\pi(3a_p^3)/m\equiv U^R_{\rm eff}=U_{\rm eff}/(1-{U_{\rm eff} \over 3}\sum_{[0,\omega_c]}{p^2 \over 2\varepsilon_{\bf p}})$, where $\omega_c$ is a high-energy cutoff. The increase of $(k_{\rm F}a_p)^{-3}$ corresponds to a decrease of bare threshold energy $2\nu$. Since the chemical potential also decreases to approach $\nu$ [see the inset in panel 2(b)], the bare interaction $U_{\rm eff}=U+g_{\rm r}^2/(2\nu-2\mu)$ becomes stronger for larger $(k_{\rm F}a_p)^{-3}$. 
In the BCS regime [$(k_{\rm F}a_p)^{-3}\lesssim -1$], 
$T_{\rm c}$ agrees well with the standard weak-coupling BCS 
theory [`BCS' in Fig. 2(a)]. 
On the other hand, in the crossover regime $[-1\lesssim (k_{\rm F}a_p)^{-3}\lesssim 0]$, the deviation of $T_{\rm c}$ from the weak-coupling result is large. The chemical potential $\mu$ also begins to strongly deviate from the Fermi energy $\varepsilon_{\rm F}$, as shown in Fig. 2(b). Figure 3 shows that the gas continuously changes from a gas of Fermi atoms (dominated by $N_{\rm F}$) into a Bose gas of bound states (dominated by $N_{\rm M}=N_{\rm B}+N_{\rm C}$). In the BEC regime [$(k_{\rm F}a_p)^{-3}\lesssim 0$], free Fermi atoms are almost absent, and $T_{\rm c}$ approaches a constant value. Its precise value depending on whether one is dealing with a single-component gas ($\uparrow\uparrow$) or a two-component gas ($\uparrow\downarrow$). This difference is due to different 
Fermi energies in the two cases (see TABLE I). 
The peak in $T_{\rm c}$ in Fig. 2(a) would be absent if the 
coupling to the bound states\cite{Haussmann,Ohashi} was properly included in the self-energies of the Fermi atoms.
\par
In the extreme BEC limit, where all the atoms have formed Feshbach molecules ($N_{\rm F},~N_{\rm C}=0$), the gas can be regarded as a non-interacting 
Bose gas mixture with three kinds of Feshbach molecules, with $L_z=\pm 1,0$. In this case, rewriting Eq. (\ref{eq.7}) as 
$2\mu=2\nu-g_{\rm r}^2\Pi_{ii}(0,0)/[1-U\Pi_{ii}(0,0)]$, we find $2\mu\to2\nu$, because $\Pi_{ii}(0,0)=0$ in this BEC limit. This result is consistent with the inset in Fig. 2(b). Since $2\mu$ is the chemical potential of the molecular Bose gas and $2\nu$ is the threshold energy of molecular excitations, the condition $2\mu=2\nu$ is that required for BEC in a non-interacting Bose gas. That is to say, $T_{\rm c}$ in this extreme case is simply determined by $N=3\sum_{\bf q}n_B(\varepsilon^B_{{\bf q}})$, where $n_B(\varepsilon)$ is the Bose distribution function. The factor 3 comes from the presence of {\it three} kinds of molecules, which is characteristic of $p$-wave superfluidity. Because of this factor, $T_{\rm c}$ in the $p$-wave case is lower than the $s$-wave case, as shown in TABLE I. TABLE I also shows $T_{\rm c}$ in the BEC limit in a trapped gas, evaluated within the LDA. These values for $T_{\rm c}$ in a trapped gas seems accessible in current experiments. The crossover behavior of $T_{\rm c}$ shown in Fig. 2(a) is a general result valid 
for any type of $p$-wave superfluidity.
\par
Figures \ref{fig2} and \ref{fig3} indicate that the crossover behavior of $T_{\rm c}$, $\mu$, and the number of Fermi atoms $N_{\rm F}$ and that of Bose molecules $N_{\rm M}$ show quasi-universal behavior when plotted as a function of $(k_{\rm F}a_s)^{-3}$, irrespective of whether the Feshbach resonance is narrow  (${\bar g}_{\rm r}<\varepsilon_{\rm F}$) or broad (${\bar g}_{\rm r}>\varepsilon_{\rm F}$). On the other hand, the character of the bound state bosons is different between the two. In a narrow Feshbach resonance, the Feshbach molecules ($N_{\rm B}$) are dominant in the crossover regime, while Cooper-pairs ($N_{\rm C}$) are dominant in a broad Feshbach resonance (see Fig. 3). However, Feshbach molecules always dominate in the extreme BEC limit. 
\par
We note that the phase diagrams in trapped Fermi gases\cite{Jin,Bartenstein,Zwierlein,Kinast,Bourdel} which experiments measure involve passing through the resonance in an adiabatic (constant entropy) manner. In the case of a $s$-wave Feshbach resonance, Ref. \cite{James} has discussed how one can calculate such phase diagrams using a simple ideal gas model. This could be extended to the $p$-wave Feshbach resonance case.
\par
To summarize, we have discussed the BCS-BEC crossover in the presence of a $p$-wave Feshbach resonance.
Generalizing earlier work on the $s$-wave BCS-BEC crossover\cite{Ohashi}, we have included fluctuation effects in the three $p$-wave Cooper-channels, as well as the three kinds of Feshbach molecules with $L_z=\pm 1,0$. Observation of the molecular condensate in the BEC regime\cite{Jin,Bartenstein,Zwierlein,Kinast,Bourdel} would be a first step in the study of $p$-wave superfluidity.
\par
I would like to thank Prof. A. Griffin for discussions and also critical reading of this manuscript. This work was financially supported by a Grant-in-Aid for Scientific research from Ministry of Education of Japan and funds from NSERC of Canada. 
%

\newpage
\begin{figure}
\caption{
\label{fig1}
(a) Particle-particle scattering matrix in the $t$-matrix approximation in terms of the
non-resonant interaction $U$ (first line) and the $p$-wave Feshbach resonance $g_{\rm r}$ (second line). $G_0$ and $D_0$ are the bare single-particle Fermi and Bose Green's function, respectively. (b) Corrections to the thermodynamic potential originating from fluctuations in the $p$-wave Cooper-channels (left diagram) and the Feshbach resonance (right diagram).
}
\end{figure}
  
\begin{figure}
\caption{
\label{fig2}
(a) $T_{\rm c}$ in the $p$-wave BCS-BEC crossover.
$\uparrow\uparrow$: single-component Fermi gas, with ${\bar U}\equiv Np_{\rm F}^2U=0.4\varepsilon_{\rm F}$. $\uparrow\downarrow$: two-component Fermi gas, with ${\bar U}=0.8\varepsilon_{\rm F}$. Results for a narrow Feshbach resonance (${\bar g}_{\rm r}\equiv\sqrt{Np^2_{\rm F}}g_{\rm r}=0.6\varepsilon_{\rm F}$) and a broad Feshbach resonance $({\bar g}_{\rm r}=5\varepsilon_{\rm F})$ are shown. `BCS' shows a weak-coupling result in the two-component case with $\mu$ being fixed at the value at $\nu=2.5\varepsilon_{\rm F}$. 
(b) Chemical potential $\mu(T_{\rm c})$ in a single component Fermi gas. The solid and dashed lines show the results for a narrow and a broad Feshbach resonance, respectively. The inset shows $\mu(T_{\rm c})$ as a function of the threshold energy $2\nu$ for a narrow Feshbach resonance.
}
\end{figure}

\begin{figure}
\caption{
\label{fig3}
Numbers for various kinds of
particles at $T_{\rm c}$ in a single-component Fermi gas. (a) narrow
 Feshbach resonance, and (b) broad Feshbach resonance. $N_{\rm M}\equiv N_{\rm B}+N_{\rm C}$, where $N_{\rm B}$ describes Feshbach molecules and $N_{\rm C}$ 
gives the contribution from Cooper-pairs (stable and unstable).
}
\end{figure}
\newpage
\centerline{}
\begin{table}
\caption
{
\label{table1}
$T_{\rm c}$ in the BEC limit. (S) and (T) show the single- and two-component Fermi gas, respectively. $T_{\rm c}$ in a uniform gas is given by $T_{\rm c}={2T_{\rm F} \over [6\alpha\sqrt{\pi}\zeta(3/2)]^{2/3}}$ (where $\zeta(z)$ is the zeta-function), 
with $\alpha=1$ ($s$-wave), $\alpha=6$ (S), and $\alpha=3$ 
(T). In a harmonic trap, $T_{\rm c}={T_{\rm F} \over [6\alpha\zeta(3)]^{1/3}}$ is evaluated using the LDA. $T_{\rm F}$ ($=\varepsilon_{\rm F}$) is obtained from 
$N=\eta\sum_{\varepsilon\le\varepsilon_{\rm F}}1$, where $\eta=1$ ($\eta=2$) for the single (two) component case.
}
\begin{ruledtabular}
\begin{tabular}{lcr}
symmetry &uniform gas [$T_{\rm F}$]&trapped gas [$T_{\rm F}$]\\
\hline
$s$-wave & 0.218 & 0.518\\
$p$-wave (S) & 0.066 & 0.285\\
$p$-wave (T) & 0.105 & 0.359\\
\end{tabular}
\end{ruledtabular}
\end{table}

%

\begin{thebibliography}{99}
\bibitem{Jin} C. Regal et al., Phys. Rev. Lett. {\bf 92}, 040403 (2004).
\bibitem{Bartenstein} M. Bartenstein et al., Phys. Rev. Lett. {\bf 92}, 120401 (2004).
\bibitem{Zwierlein} M. Zwierlein et al., Phys. Rev. Lett. {\bf 92}, 120403 (2004).
\bibitem{Kinast} J. Kinast et al., Phys. Rev. Lett. {\bf 92}, 150402 (2004).
\bibitem{Bourdel} T. Bourdel et al., Phys. Rev. Lett. {\bf 91}, 020402 (2004).
\bibitem{Regal} C. Regal et al., Phys. Rev. Lett. {\bf 90}, 053201 (2003).
\bibitem{Salomon} J. Zhang, et al., cond-mat/0406085.
\bibitem{Schunk} C. Schunk, et al., cond-mat/0407373.
\bibitem{Nozieres} P. Nozi\`eres and S. Schmitt-Rink, J. Low. Temp. Phys. {\bf 59}, 195 (1985).
\bibitem{Ohashi} Y. Ohashi and A. Griffin, Phys. Rev. Lett. {\bf 89}, 130402 (2002). See also Y. Ohashi and A. Griffin, Phys. Rev. A {\bf 67}, 033603 (2003); {\bf 67}, 063612 (2003).
\bibitem{Stoof} M. Houbiers, et al., Phys. Rev. A {\bf 56}, 4864 (1997).
\bibitem{Bohn} J. Bohn, Phys. Rev. A {\bf 61}, 053409 (2000).
\bibitem{Ho} T. Ho and R. Diener, cond-mat/0408468.
\bibitem{Melo} S. Botelho and C. S\'a de Melo, cond-mat/0409357.
\bibitem{Baranov} M. Baranov et al., cond-mat/0409150.
\bibitem{Timmermans} E. Timmermans et al., Phys. Lett. A {\bf 285}, 228 (2001).
\bibitem{Holland} M. Holland et al., Phys. Rev. Lett. {\bf 87}, 120406 (2001).
\bibitem{Voll} D. Vollhardt and P. W\"olfle, in {\it The Superfluid Phases of He3} (Taylor \& Francis, N.Y., 1990), Chap. 3.
\bibitem{Eagles} D. Eagles, Phys. Rev. {\bf 186}, 456 (1969).
\bibitem{Leggett} A. Leggett, in {\it Modern Trends in the Theory of
    Condensed Matter}, edited by A. Pekalski and J. Przystawa (Springer
  Verlag, Berlin, 1980), p. 14.
\bibitem{Ohashi2} Y. Ohashi and A. Griffin, cond-mat/0410220, Sec. IV.
\bibitem{Haussmann} R. Haussmann, Phys. Rev. B. {\bf 49}, 12975 (1994).
\bibitem{James} J. Williams et al., New J. Phys. {\bf 6}, 123 (2004).
\end{thebibliography}
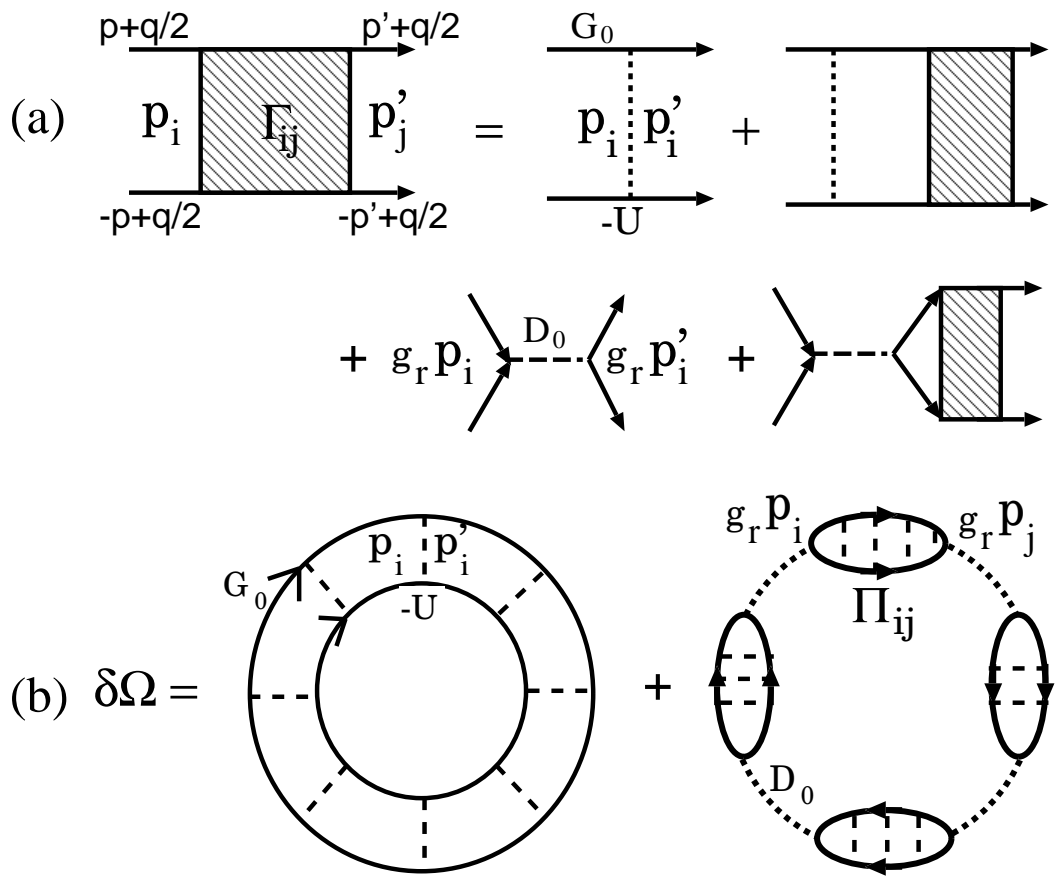
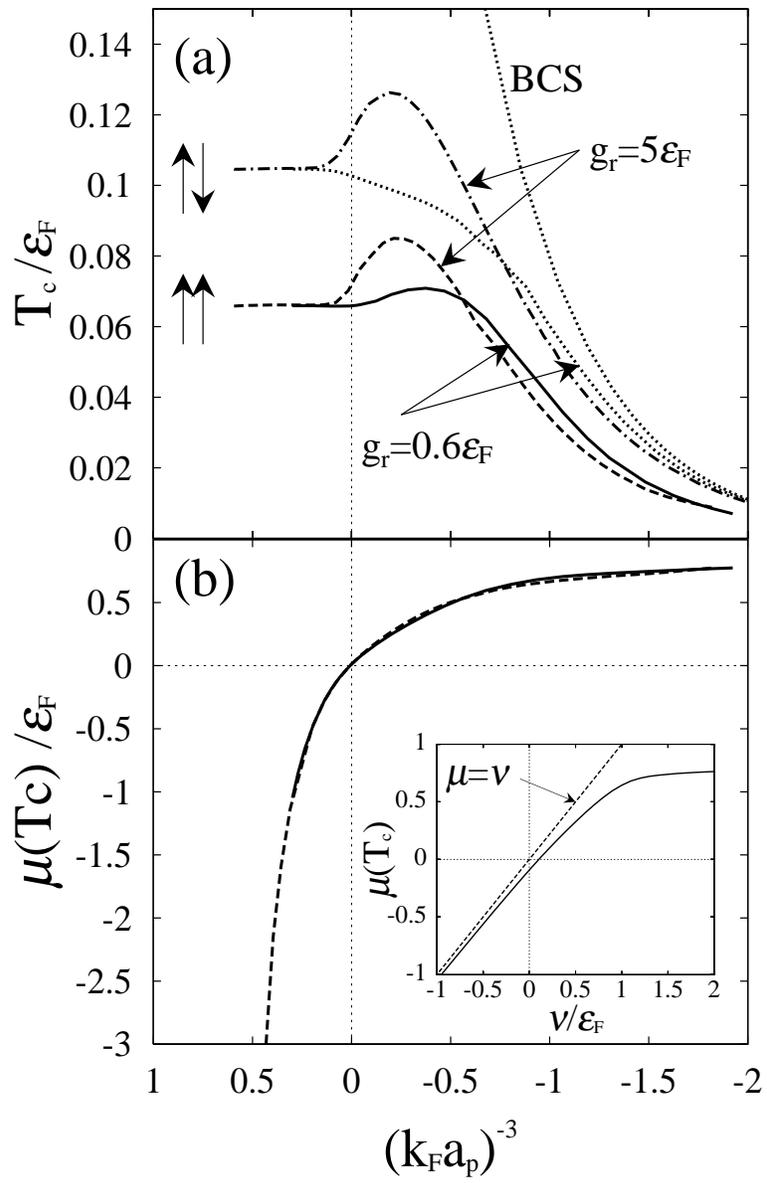
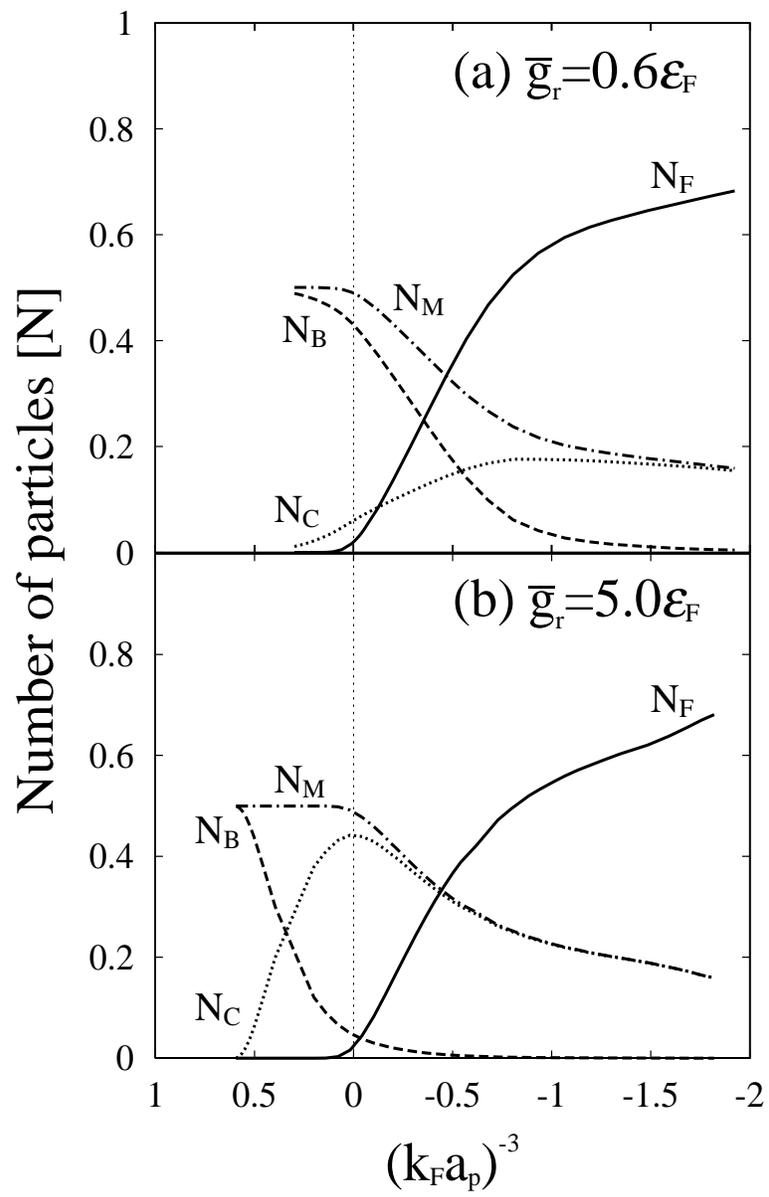
\end{document}